# Ruin models with investment income

## Jostein Paulsen*

*Department of Mathematics*
*University of Bergen*
*Johs. Brunsgt. 12*
*5008 Bergen, Norway*

**Abstract:** This survey treats the problem of ruin in a risk model when assets earn investment income. In addition to a general presentation of the problem, topics covered are a presentation of the relevant integro-differential equations, exact and numerical solutions, asymptotic results, bounds on the ruin probability and also the possibility of minimizing the ruin probability by investment and possibly reinsurance control. The main emphasis is on continuous time models, but discrete time models are also covered. A fairly extensive list of references is provided, particularly of papers published after 1998. For more references to papers published before that, the reader can consult [47].

**AMS 2000 subject classifications:** Primary 60G99; secondary 60G40, 60G44, 60J25, 60J75.

**Keywords and phrases:** Ruin probability, Risk theory, Compounding assets.

## Contents

## 1. Introduction

The problem of ruin has a long history in risk theory, going back to Lundberg [40]. In Lundberg's model, the company did not earn any investment on its capital. An obvious reason for this assumption, although there may be other reasons as well, is that the mathematics is easier, and back in the first half of

---

the 20th century the theory of stochastic processes was far less developed, and also far less known, than it is today. The first attempt to incorporate investment incomes was undertaken by Segerdahl in [62]. Segerdahls assumption was that capital earns interest at a fixed rate $r$. This model was further elaborated in [22, 28], and in a somewhat more general form in [18]. The books [4, 54] both have sections devoted to it, and it remains very popular even today. We will repeatedly return to it in this survey.

Inspired by ideas from mathematical finance, in [45] a model was suggested where capital is allowed to be invested in risky assets. Our starting point here will be a slightly restricted version of this model.

For a survey of the theory before 1998, the reader is referred to [47]. Since 1998, there are particularly three new developments that has influenced and given new vitality to this topic.

1. The emphasis on heavy tailed claim distributions.
2. The Gerber-Shiu penalty function.
3. The possibility to influence the ruin probability by control of the risky investments and possibly reinsurance.

In 1998, the only papers dealing with these items in the context of this survey were [3, 7, 35].

In order not to be bogged down in technicalities, the reader is referred to the references for special cases and detailed assumptions.

To make the ideas transparent, we introduce the risk process by means of two basic processes, i.e.

- A basic risk process $P$ with $P_0 = 0$.
- A return on investment generating process $R$ with $R_0 = 0$.

It is assumed throughout in this survey that $P$ and $R$ are independent. If $P$ and $R$ belong to the rather general class of semimartingales, then under some weak additional assumptions we can define the risk process as

$$Y_t = y + P_t + \int_0^t Y_{s-} dR_s, \qquad (1.1)$$

so that $Y_0 = y$. By [47] the solution of this equation is given as

$$Y_t = e^{\tilde{R}_t} \left( y + \int_0^t e^{-\tilde{R}_s} dP_s \right),$$

where $\tilde{R} = \log \mathcal{E}(R)$ is the logarithm of the Doléans-Dade exponential of $R$, i.e. $V_t = \mathcal{E}(R)_t$, satisfies the stochastic differential equation $dV_t = V_{t-} dR_t$ and $V_0 = 1$.

In this survey we shall typically assume that $P$ and $R$ are of the forms

$$P_t = pt + \sigma_P W_{P,t} - \sum_{i=1}^{N_t} S_i, \qquad (1.2)$$

$$R_t = rt + \sigma_R W_{R,t}, \qquad (1.3)$$

where $W_P$ and $W_R$ are Brownian motions, $N$ a Poisson process with rate $\lambda$ and the $\{S_i\}$ are positive, independent and identically distributed random variables (i.i.d.) with distribution function $F$. Furthermore, $W_P$, $W_R$, $N$ the $\{S_i\}$ are all independent. The idea is that $p$ is the premium rate, the $\{S_i\}$ are claims while $W_P$ represents fluctuations in premium income and maybe also small claims. The return on investment generating process $R$ is the standard Black Scholes return process. With these assumptions, $Y$ becomes a homogeneous, strong Markov process, a fact that allows us to draw on the vast literature on Markov processes. When $R$ follows (1.3), $\tilde{R}_t = R_t - \frac{1}{2}\sigma_R^2 t = (r - \frac{1}{2}\sigma_R^2)t + \sigma_R W_{R,t}$.

A few papers [45, 49, 75, 76] generalize the return on investment process $R$ to the jump-diffusion

$$R_t = rt + \sigma_R W_{R,t} + \sum_{i=1}^{N_{R,t}} S_{R,i}, \qquad (1.4)$$

where $\sum_{i=1}^{N_{R,t}} S_{R,i}$ is another independent compound Poisson process. Letting $F_R(x) = P(S_R \leq x)$, in order to avoid certain ruin caused by losing everything in an investment shock, it is assumed that $F_R(-1) = 0$. At an even higher level of generality, in [46, 48], $P$ and $R$ are independent Lévy processes.

The time of ruin is defined as $T = \inf\{t : Y_t < 0\}$, with $T = \infty$ if $Y$ stays positive. The probability of ruin in finite versus infinite time is then defined as

$$\psi(t,y) = P(T \leq t | Y_0 = y) \quad \text{and} \quad \psi(y) = P(T < \infty | Y_0 = y).$$

Mathematically, $\psi(y)$ is the easiest, and probably as a consequence it has by far been the most popular in the literature.

A more general concept that also allows for problems relating to the time of ruin, the size of the deficit at ruin or the surplus immediately before ruin, is the Gerber-Shiu penalty function [23]. It is given as

$$\Phi_\alpha(y) = E[g(Y_{T-}, |Y_T|)e^{-\alpha T} 1_{\{T<\infty\}} | Y_0 = y],$$

where $g$ is a nonnegative function and $\alpha \geq 0$. Various choices of $g$ and $\alpha$ allow for the computation of many interesting quantities related to ruin [9]. Note that with $g(y_1, y_2) = 1$ and $\alpha = 0$, $\Phi_0(y) = \psi(y)$. Another choice is $g(y_1, y_2) = 1_{\{y_1 \leq u, y_2 \leq v\}}$ and $\alpha = 0$ giving

$$H(y, u, v) = P(T < \infty, Y_{T-} \leq u, |Y_T| \leq v | Y_0 = y). \qquad (1.5)$$

Papers with material on $\Phi_\alpha(y)$ or special cases of it are [9, 12, 70, 71, 72, 73, 74, 75, 76, 77].

A somewhat different ruin concept introduced in [19] and [57] is the model for absolute ruin. Let $\bar{y} \geq 0$ and

$$Y_t = y + P_t + \int_0^t Y_{s-} 1_{\{Y_{s-} < 0\}} dR_{1,s} + \int_0^t (Y_{s-} - \bar{y}) 1_{\{Y_{s-} > \bar{y}\}} dR_{3,s}, \qquad (1.6)$$

where $R_{1,t} = r_1 t$ and $R_{3,t} = r_3 t + \sigma_R W_{R,t}$. The idea is that when capital is negative, money can be borrowed at rate $r_1$. When capital is between 0 and $\bar{y}$, it is kept in the company without earning any investment income, while excess capital over $\bar{y}$ are invested in a possibly risky market. If $P$ follows (1.2) with $\sigma_P = 0$ and for some $T^A$, $Y_{T^A} \leq -p/r_1$, premium income is no larger than interest on debt, and consequently $Y$ drifts towards minus infinity. The time $T^A$ is called the time of absolute ruin, and $\psi^A(y) = P(T^A < \infty)$ the probability of absolute ruin. When $\sigma_P > 0$ the concept of absolute ruin is less meaningful since whatever small $Y_t$ is, it can with some positive probability drift back into positive values again.

## 2. Some general results in the Markov model

Unless otherwise stated, it is assumed that $P$ and $R$ follow (1.2) and (1.3). Then it follows from [46, 52] that under weak assumptions, $\psi(y) < 1$ if $r > \frac{1}{2}\sigma_R^2$ (and in fact when $\sigma_R^2 > 0$ then $\psi(y) < 1$ if and only if $r > \frac{1}{2}\sigma_R^2$). In this case, under weak assumptions, $\psi$ is twice continuously differentiable on $(0, \infty)$ and is a solution of the equation, see [25, 49, 69] and in particular [26],

$$L\psi(y) = -\lambda \bar{F}(y), \tag{2.1}$$

with boundary conditions

$$\lim_{y \to \infty} \psi(y) = 0 \quad \text{and} \quad \psi(0) = 1 \text{ if } \sigma_P > 0.$$

Here $L$ is the integro-differential operator

$$Lh(y) = \frac{1}{2}(\sigma_P^2 + \sigma_R^2 y^2)h''(y) + (p + ry)h'(y) + \lambda \int_0^y h(y-x)dF(x) - \lambda h(y),$$

and $\bar{F}(y) = 1 - F(y)$. Sometimes it is more convenient to work with the survival probability $\phi(y) = 1 - \psi(y)$, in which case (2.1) becomes

$$L\phi(y) = 0. \tag{2.2}$$

When $R$ instead follows (1.4), it was shown in [75] that under rather strong assumptions, $\psi$ is twice continuously differentiable and satisfies $L_R \psi(y) = -\lambda \bar{F}(y)$, where

$$L_R h(y) = Lh(y) + \lambda_R \int_{-1}^{\infty} h(y(1+x))dF_R(x) - \lambda_R h(y).$$

Equations similar to (2.1) also hold for other relevant quantities. An example is the decomposition of the ruin probability into $\psi(y) = \psi_d(y) + \psi_s(y)$, where $\psi_d(y)$ is the probability that $Y$ will become negative due to a drift in $W_P$, while $\psi_s(y)$ is the same, but due to a claim, i.e. a jump. Under suitable differentiability assumptions, it is shown in [11] that $\psi_d$ satisfies $L\psi_d(y) = 0$ with the same boundary conditions as $\psi$, while $\psi_s$ satisfies (2.1), but with $\psi_s(0) = 0$.

A further example is the calculation of the Gerber-Shiu penalty function. With $\sigma_P = 0$ and some rather strong assumptions on the distribution $F$ and the function $g$, it is shown in [9] that $\Phi_\alpha(y)$ satisfies

$$L\Phi_\alpha(y) - \alpha\Phi_\alpha(y) = -\lambda \int_y^\infty g(y, x-y) dF(x), \tag{2.3}$$

with boundary condition $\lim_{y\to\infty} \Phi_\alpha(y) = 0$. The extension to $\sigma_P > 0$ seems rather straightforward, in which case the boundary conditon $\Phi_\alpha(0) = g(0,0)$ must be added. Also, using methods such as in [26], the assumptions can probably be weakened considerably.

In the absolute ruin problem there are three equations $L_i \psi_i^A(y) = -\lambda \bar{F}(y)$, where the operators $L_i$ are as $L$ above, but referring to the different $R_i$ where $R_{2,t} = 0$. The solution can then be found by imposing proper boundary and continuity conditions.

Following ideas from [28], it was shown in [45] that the ruin probability can be written as

$$\psi(y) = \frac{H(-y)}{E[H(-Y_T)|T < \infty]}, \tag{2.4}$$

where $H$ is the distribution function of the perpetuity

$$X = \int_0^\infty e^{-(r - \frac{1}{2}\sigma_R^2)t + \sigma_R W_{R,t}} dP_t.$$

Actually, (2.4) holds whenever $P$ and $R$ are independent Lévy processes, see [46].

Corresponding to (2.1), the finite time ruin probability $\psi(t, y)$ should be the solution of the partial integro-differential equation

$$-\frac{\partial}{\partial t}\psi(t,y) + L\psi(t,y) = -\lambda \bar{F}(y), \tag{2.5}$$

with the additional boundary condition $\psi(0, y) = 1_{\{y<0\}}$. Here the operator $L$ acts on the $y$ variable. When $\sigma_P = \sigma_R = 0$ standard methods can be used to prove that the ruin probability is sufficiently smooth and satisfies (2.5), but as for the infinite time case this problem is much more complicated when $\sigma_P^2 + \sigma_R^2 > 0$, and for this case there are no results as of the existence of a smooth solution of (2.5). In [42] an equation similar to (2.5) for diffusions is discussed.

## 3. A discrete time model

With $\{\tau_i\}$ the jump-times of $P$ and $\tau_0 = 0$, setting $X_n = Y_{\tau_n}$ yields

$$X_n = e^{\tilde{R}_{\tau_n} - \tilde{R}_{\tau_{n-1}}} \left( X_{n-1} + \int_{\tau_{n-1}}^{\tau_n} e^{-(\tilde{R}_s - \tilde{R}_{\tau_{n-1}})} dP_s \right) = \xi_n X_{n-1} + \eta_n. \tag{3.1}$$

Here the sequence $\{(\xi_n, \eta_n)\}$ is i.i.d., but $\xi_n$ and $\eta_n$ are themselves not independent. However, $(\xi_n, \eta_n)$ is independent of $X_{n-1}$.

Assume that $\sigma_P = 0$, in which case

$$\psi(y) = P(X_n < 0 \text{ for some } n),$$

hence we can alternatively focus on the sequence $\{X_n\}$. An advantage with this setup is that the claim-number process $N$ does not have to be a Poisson process, any renewal process will do, [8, 77]. Actually, (3.1) can serve as a link between the time continuous and the time discrete frameworks. To see how, consider the time discrete process $\{X_n\}$ where $X_n$ is total capital at time $n$. Let $\{(\bar{R}_n, \bar{P}_n)\}$ be i.i.d. where $\bar{R}_n$ is the interest rate between time $n-1$ and $n$, and $\bar{P}_n$ is premium minus claims in the same time interval. Then

$$X_n = (1 + \bar{R}_n)X_{n-1} + \bar{P}_n = \xi_n X_{n-1} + \eta_n \quad (3.2)$$

with $\xi_n = 1 + \bar{R}_n$ and $\eta_n = \bar{P}_n$. A major difference between the time-continuous and the time-discrete setup is that in the time-discrete case it is common to assume that $\bar{R}_n$ and $\bar{P}_n$ are independent, i.e. that $\xi_n$ and $\eta_n$ are independent. The rationale for this assumption is that premiums earned and claims paid do not influence investment income until the beginning of the next period. This is the assumption in [43, 64, 65], but not in [44] where a more general setup was used in order to deal with the continuous time case as well, including when $\sigma_P \neq 0$. Also, in [16], the independence assumption of the $\xi_n$ is relaxed. In [74] a discrete time version of (2.4) is given, while [16, 43, 44, 64, 65] all focus on the asymptotics of ruin probabilities, to be briefly reviewed below.

## 4. Analytical and numerical solutions

Since most analytical results are for the infinite time horizon, we start with this case. When $\lambda = 0$ the process is a pure diffusion process and the problem is solvable. A solution can be found in [45] or [55]. When $\lambda > 0$ analytical solutions can only be obtained in a few rather simple cases, and most of these solutions are very complex. In all known solutions it is assumed that $\sigma_R = 0$, so in the sequel we shall therefore tacitly let $\sigma_R = 0$.

We have already mentioned Segerdahl's classical work when $\sigma_P = 0$ and exponentially distributed claims. In [49] these solutions were extended to the case when claims are mixtures of two exponential distributions, as well as to the case when they are Erlang(2) distributed, i.e. they can be represented as a sum of two independent exponentials. Extensions beyond that seem very difficult though. In [49] the case with $\sigma_P > 0$ and claims exponentially distributed was also solved, and this solution was extended in [10] with separate solutions for $\psi_d(y)$ and $\psi_s(y)$.

Again with $\sigma_R = 0$, the absolute ruin problem when $\sigma_P = 0$ and exponential claims was solved in [19]. This result was extended to $\sigma_P > 0$ and $\bar{y} = 0$ in [24]. Another extension can be found in [12] where rather explicit expressions for the Gerber-Shiu penalty function are given for the case with $\bar{y} = \infty$ and $\sigma_P = 0$ and claims exponentially distributed. Similar results for the case (1.5) are found in [73].

In the finite time horizon case analytical solutions are hard to come by. Using (2.5), in [36] the Laplace transform of the survival probability $\phi(t,y) = 1 - \psi(t,y)$ is found when $\sigma_P = \sigma_R = 0$ and claims are exponentially distributed. However, this transform involves a ratio of confluent hypergeometric functions, and is therefore difficult to invert, the exception is when $\lambda = r$ in which case the solution is rather simple. A different method for the same problem is used in [1] who provide a recursion for $\phi(t,y)$ when $\lambda = kr$ for some positive integer $k$. This recursion is solved and exact solutions given for $k=1$ and $k=2$.

The value of continuous time ruin theory is mostly due to its simplicity as a concept together with its complexity as a mathematical problem, and this can explain why the issue of computing numerical values has received comparatively little attention. In [51], following an idea from [63], but allowing $\sigma_P^2 > 0$, using integration by parts the equation (2.2) was turned into a Volterra integral equation and methods from numerical analysis was used to solve this numerically. In the finite time case several methods have been proposed when $\sigma_P = \sigma_R = 0$, see e.g. [6, 13], and for a somewhat more general discrete time model [17]. These methods are rather intuitive in nature and not based on any particular known procedure from numerical analysis. Their efficiency are therefore somewhat low, but as a bonus they provide upper and lower bounds.

An alternative to traditional numerical methods is Monte Carlo simulation, which is particularly well suited for finite time ruin problems. For infinite time ruin problems some care has to be taken as to when the simulation should stop. An alternative is to simulate under an equivalent measure $\tilde{P}$ so that $\tilde{P}(T < \infty | Y_0 = y) = 1$. Then with $M_t$ equal to $\frac{d\tilde{P}}{dP}$ restricted to $\sigma\{Y_s : s \leq t\}$,

$$\psi(y) = \tilde{E}[M_T^{-1} | Y_0 = y],$$

hence what is required is repetitive simulation of $M_T$ under $\tilde{P}$. The catch is that this kind of importance sampling is very dangerous, and can lead to completely wrong results, see e.g. [50] for examples. In [5] a somewhat complicated but efficient change of measure is given for the case with $\sigma_P = \sigma_R = 0$ and light-tailed claims. The change of measure method also works for finite time ruin probabilities, in which case $\psi(t,y) = \tilde{E}[M_{T \wedge t}^{-1} | Y_0 = y]$.

## 5. Asymptotic results

If there is only a few papers dealing with numerical solutions, there are certainly a fair amount dealing with asymptotic results, i.e. the behaviour of $\psi(y)$ or $\psi(t,y)$ when $y$ gets large. To present a few of the most prominent or recent results we shall need some definitions. As before let $\bar{F}(x) = 1 - F(x)$ and let $F^{*n}(x)$ be the $n$-fold convolution of $F$.

- $F \in \mathcal{R}_{-\alpha}$ if $\bar{F}(x) = x^{-\alpha} l(x)$ where $l$ is a slowly varying function, i.e. $\lim_{x \to \infty} \frac{l(tx)}{l(x)} = 1$ for all $t > 0$.

- $F \in \mathrm{ERV}(\alpha, \beta)$ if for $0 < \alpha \leq \beta < \infty$ and all $t > 1$,

$$t^{-\beta} \leq \liminf_{x \to \infty} \frac{\bar{F}(tx)}{\bar{F}(x)} \leq \limsup_{x \to \infty} \frac{\bar{F}(tx)}{\bar{F}(x)} \leq t^{-\alpha}.$$

- $F \in \mathcal{S}$ if $\lim_{x \to \infty} \frac{\overline{F^{*2}}(x)}{\bar{F}(x)} = 2$.

Clearly $\mathcal{R}_{-\alpha} = \mathrm{ERV}(\alpha, \alpha)$. We just write $F \in \mathrm{ERV}$ if $F \in \mathrm{ERV}(\alpha, \beta)$ for some $0 < \alpha \leq \beta$. It can be proved that $\mathrm{ERV} \subset \mathcal{S}$. The class $\mathcal{S}$ is called the class of subexponential distributions, and among others it contains the lognormal, loggamma as well as the heavy tailed Weibull distribution, i.e. $\bar{F}(x) = e^{-(x/\beta)^\gamma}$ with $\gamma < 1$. It can be shown that if $F \in \mathcal{S}$ then $E[e^{tS}] = \infty$ for all $t > 0$, hence the name subexponential distribution.

Culminating through a series of papers [3, 33, 35, 37, 66, 67, 68], it was proved in [32] that when $\sigma_R = 0$ and $F \in \mathcal{S}$,

$$\psi(t, y) \sim \frac{\lambda}{r} \int_y^{ye^{rt}} \frac{\bar{F}(x)}{x} dx, \quad 0 < t \leq \infty. \tag{5.1}$$

In addition, in the same paper it was proved that if $F \in \mathcal{R}_{-\alpha}$ for some $\alpha > 0$ and $N$ is a renewal process independent of the $S_i$ and $W_P$,

$$\psi(t, y) \sim \bar{F}(y) \int_0^t e^{-\alpha rs} dm(s), \tag{5.2}$$

where $m$ is the renewal measure of $N$. In particular, if $N$ is the Poisson process then $m(s) = \lambda s$, hence

$$\psi(t, y) \sim \frac{\lambda}{\alpha r} \bar{F}(y)(1 - e^{-\alpha rt}), \quad 0 < t \leq \infty.$$

A generalization of (5.2) where it is only assumed that $F \in \mathrm{ERV}$ is given in [15] when $t = \infty$. Under the additional weakened assumption that claims are pairwise negative dependent, i.e. for $i \neq j$,

$$P(S_i \leq u, S_j \leq v) \leq F(u)F(v),$$

they show that

$$\psi(y) \sim \int_0^\infty \bar{F}(ye^{rs}) dm(s),$$

which is easily seen to be the same as (5.2) when $F \in \mathcal{R}_{-\alpha}$. Another generalization is provided in [78], where under some extra conditions the formula (5.1) is proved to hold for the absolute ruin problem when $t = \infty$, $\sigma_P = 0$ and $\bar{y} = 0$ in (1.6).

For the light tailed case with $\sigma_R = 0$ results are a little less explicit. Let $\kappa = \sup\{a : E[e^{aS}] < \infty\}$. Then it was proved in [19] and [57] that for $\sigma_P = 0$ and any $\varepsilon > 0$,

$$\lim_{y \to \infty} e^{(\kappa - \varepsilon)y} \psi(y) = 0 \quad \text{and} \quad \lim_{y \to \infty} e^{(\kappa + \varepsilon)y} \psi(y) = \infty. \tag{5.3}$$

It was also shown by examples that anything can happen to $\lim_{y\to\infty} e^{\kappa y}\psi(y)$. The result (5.3) was also proved for the absolute ruin problem with $\bar{y} = 0$, and a simplified proof can be found in [78].

When $\sigma_R > 0$ the picture is somewhat less complex. The reason for this is that the financial risk caused by variations in the return processs $R$ corresponds to claims $F \in \mathcal{R}_{-\rho}$ with $\rho = \frac{2r}{\sigma_R^2} - 1$. So when claims have lighter tails than this, the asymptotics is dominated by $R$. Various results in this direction appear in [20, 34, 44]. The most precise results for the model studied here are those in [26]. There it is assumed that $\sigma_P = 0$, but due to the light tailed effect of $W_P$, the result is undoubtedly valid for $\sigma_P > 0$ as well. To present the results, remember from the beginning of Section 2 that $\psi(y) = 1$ when $\rho \le 0$, so assume that $\rho > 0$ and that $E[S] < \infty$.

1. If for some $\varepsilon > 0$, $E[S^{\rho+\varepsilon}] < \infty$, then $\lim_{y\to\infty} y^\rho \psi(y) = c$ for some (for all practical purposes) unknown $c$.
2. If $F \in \mathcal{R}_{-\alpha}$ where $\alpha < \rho$, then

$$\psi(y) \sim \frac{2\lambda}{\sigma_R^2 \alpha(\rho - \alpha)} \bar{F}(y).$$

3. If $F \in \mathcal{R}_{-\rho}$ with $\bar{F}(x) = x^{-\rho} l(x)$ and $M < \infty$ where $M = \int_1^\infty \frac{l(x)}{x} dx$, then $\psi(y)$ is asymptotically as in Case 1 above. If $M = \infty$ then

$$\psi(y) \sim \frac{2\lambda}{\sigma_R^2 \rho} y^{-\rho} \int_1^y \frac{l(x)}{x} dx.$$

In [2] similar results, but somewhat less explicit, are proved when claim intervals are either Erlang distributed or a mixture of two exponential distributions. Asymptotics when $P$ and $R$ are independent Lévy processes can be found in [48].

There is a considerable literature devoted to the asymptotics of (3.2). In [43, 44] asymptotics for $\psi(y)$ and $\psi(n \log y, n)$ for $n$ fixed is studied, mostly in a large deviations context. In [64, 65] asymptotics for $\psi(n, y)$ with $n$ fixed and $y$ increasing is provided. This is done for a variety of different assumptions on the tail behaviour of the claims, and their results are similar to those reported above for $\psi(y)$ in the time continuous case. A generalization of the asymptotics for $\psi(y)$, allowing for dependence in the $\{\xi_n\}$ process is provided in [16], where it is shown that $\psi(y)$ decreases at a power rate under rather general Markovian assumptions.

A slightly different kind of result in the time discrete case is in [64, 65] and generalized in [14] where it is proved that when $F \in \mathcal{S}$, under some rather mild conditions

$$\psi(n, y) \sim \sum_{k=1}^n P\left(\eta \prod_{i=1}^k \xi_i < -y\right),$$

where $\eta$ and the $\xi_i$ are all independent and defined as in (3.2). For more asymptotic results the reader should consult the references.

## 6. Inequalities

Several inequalities for the ruin probability exist, but the problem is that either the bounds are too conservative, or that they are difficult to compute. An example that follows directly from (2.4) is

$$H(-y) \leq \psi(y) \leq \frac{H(-y)}{H(0)},$$

and when $F$ has a decreasing failure rate, the right side can be strengthened to $\frac{H(-y)}{E[H(S)]}$. The problem here is of course that the distribution function $H$ is not known.

An upper bound that is easier to compute, but restricted to the case with $\sigma_P = \sigma_R = 0$ and light tailed claims is given in [5]. To explain, assume that the equation

$$\lambda \left( E[e^{\gamma(u)S}] - 1 \right) - \gamma(u)(p + ru) = 0$$

has a positive solution $\gamma^*(u)$ for all $0 \leq u \leq y$. Then

$$\psi(y) \leq e^{-\int_0^y \gamma^*(u)du}.$$

Further examples of inequalities can be found in [8, 34, 37, 38, 41, 75].

## 7. Stochastic interest rates

In this section the model is slightly changed. The risk process $P$ is still given by (1.3), but instead of investing in risky assets, surplus capital is invested in a money account. Letting $r_t$ be the rate of return of this account, capital at time $t$ now becomes

$$Y_t = y + P_t + \int_0^t r_s Y_s ds.$$

Assume that $r$ follows the stochastic differential equation

$$dr_t = \mu(r_t)dt + \sigma(r_t)dW_{R,t},$$

where $\mu$ and $\sigma$ satisfy the necessary conditions for existence and uniqueness, and where again $W_R$ is a Brownian motion independent of $W_P$. Then $Y$ is no longer a Markov process, but the joint process $(r, Y)$ is a two-dimensional Markov process. Therefore, the probability of ruin will depend on the state of both these factors, and to avoid confusion with earlier notation we write $\tilde{\psi}(r, y)$ for the infinite time ruin probability. Forgetting about existence and smoothness properties, straightforward use of Itô's formula gives that $\tilde{\psi}(r, y)$ satisfies the following partial integro-differential equation

$$\frac{1}{2}\sigma^2(r)\tilde{\psi}_{rr}(r,y) + \frac{1}{2}\sigma_P^2\tilde{\psi}_{yy}(r,y) + \mu(r)\tilde{\psi}_r(r,y) + (ry + p)\tilde{\psi}_y(r,y)$$
$$+ \lambda \int_0^y \tilde{\psi}(r, y - x)dF(x) - \lambda\tilde{\psi}(r, y) = -\lambda\bar{F}(y). \tag{7.1}$$

Here $\tilde{\psi}_r(r,y) = \frac{\partial}{\partial r}\tilde{\psi}(r,y)$, $\tilde{\psi}_{rr}(r,y) = \frac{\partial^2}{\partial r^2}\tilde{\psi}(r,y)$ and so on. In addition appropriate boundary conditions are needed.

Needless to say, this is a very complicated equation, even in the pure diffusion case when $\lambda = 0$. Assuming that $\lambda = 0$ and that $r$ follows the simple Vasicek model

$$dr_t = (a - br_t)dt + \sigma_R dW_{R,t},$$

(7.1) becomes the elliptic PDE

$$\frac{1}{2}\sigma_R^2 \tilde{\psi}_{rr}(r,y) + \frac{1}{2}\sigma_P^2 \tilde{\psi}_{yy}(r,y) + (a-br)\tilde{\psi}_r(r,y) + (ry+p)\tilde{\psi}_y(r,y) = 0,$$

on $(-\infty, \infty) \times (0, \infty)$. Boundary conditons are

$$\begin{aligned}
\tilde{\psi}(r,0) &= 1, \\
\lim_{r \to -\infty} \tilde{\psi}(r,y) &= 1, \\
\lim_{r \to \infty} \tilde{\psi}(r,y) &= 0, \quad y > 0, \\
\lim_{y \to \infty} \tilde{\psi}(r,y) &= 0.
\end{aligned}$$

At the time of writing no results are known for this kind of problem.

## 8. Minimization of ruin probabilities

Returning to the basic model (1.1)–(1.3), we will now assume that in additon to investing in the risky asset the company can also invest in a risk free asset with return $r_f$, where $r_f < r$. We denote by $a_t$ the proportion of the capital invested in the risky asset, so that $1 - a_t$ is the proportion invested in the risk free asset. With this, the analogue to (1.3) is

$$dR_t^a = ((1-a_t)r_f + a_t r)dt + a_t \sigma_R dW_{R,t} = (r_f + a_t(r-r_f))dt + a_t \sigma_R dW_{R,t}. \quad (8.1)$$

Clearly, if $a_t \equiv a$, a constant, then (1.3) and (8.1) are equivalent.

In addition we may allow proportional reinsurance. Let $b_t$ be the fraction of the risk retained in the company. Then the fraction $1 - b_t$ is covered by the reinsurers, and we assume that the premium rate for this reinsurance is $(1 - b_t)\eta p$, where $\eta > 1$. Thus $\eta - 1 > 0$ is an additional charge made by the reinsurers. This may not always be realistic, but for the problem here it is necessary in order to avoid trivial solutions. With this, the analogue of (1.2) is

$$\begin{aligned}
dP_t^b &= p(1 - (1-b_t)\eta)dt + b_t \sigma_P dW_{P,t} - b_t d\left(\sum_{i=1}^{N_t} S_i\right) \\
&= b_t dP_t - (1-b_t)(\eta - 1)pdt.
\end{aligned} \quad (8.2)$$

Again if $b_t \equiv b$, a constant, then (1.2) and (8.2) are equivalent.

In order to be admissible, the controls $a_t$ and $b_t$ must belong to certain sets $\mathcal{A}$ and $\mathcal{B}$ respectively. Examples for the set $\mathcal{A}$ can be:

| $\mathcal{A}$ | Restrictions |
|---|---|
| $(-\infty, \infty)$ | No restrictions |
| $(-\infty, 1]$ | Short sale allowed, but no borrowing |
| $[0, \infty)$ | No short sale, but borrowing is allowed |
| $[0, 1]$ | No short sale and no borrowing |

There are of course other possibilities as well, and with a similar set of possibilities for the set $\mathcal{B}$, we see that the number of interesting cases is very high, and dealing with all of them is not practical. In practice the most interesting case is $\mathcal{A} \times \mathcal{B} = [0, 1] \times [0, 1]$. However, since it is simpler, but also since it is useful to see how much can be gained without restrictions, most results quoted here are for the case $\mathcal{A} \times \mathcal{B} = (-\infty, \infty) \times (-\infty, \infty)$, or the case $\mathcal{A} \times \mathcal{B} = (-\infty, \infty) \times \{1\}$ when there are no reinsurance options.

As in (1.1) we can now let

$$Y_t^{a,b} = y + P_t^b + \int_0^t Y_{s-}^{a,b} dR_s^a,$$

and $\psi^{a,b}(y) = P(T^{a,b} < \infty)$ where $T^{a,b} = \inf\{t : Y_t^{a,b} < 0\}$. Then define

$$\psi^{a^*, b^*}(y) = \min_{a \in \mathcal{A}, b \in \mathcal{B}} \psi^{a,b}(y),$$

whenever the minimum exists. If $\psi(y)$ is the ruin probability without control options, then clearly $\psi^{a^*, b^*}(y) \leq \psi(y)$. If they exist, the minimizing policies will be denoted by $a_t^*$ and $b_t^*$. Sometimes it is more practical to consider total amount invested in the risky asset, so we write $A_t = a_t Y_t^{a,b}$. Since optimal controls are Markov controls, we can write $a_t^* = a^*(Y_t^{a^*, b^*})$ and so on. If there are no reinsurance options, i.e. $\mathcal{B} = \{1\}$, we just write $\psi^a(y)$.

An equivalent problem is of course to maximize the survival probability $\phi^{a,b}(y) = 1 - \psi^{a,b}(y)$. The Hamilton-Jacobi-Bellman (HJB) equation for this problem is

$$\max_{a \in \mathcal{A}, b \in \mathcal{B}} \left\{ ((r_f + a(r - r_f))y + p(1 - (1-b)\eta)) \frac{d}{dy} \phi^{a,b}(y) \right.$$

$$\left. + \frac{1}{2}(a^2 \sigma_R^2 y^2 + b^2 \sigma_P^2) \frac{d^2}{dy^2} \phi^{a,b}(y) + \lambda(E[\phi^{a,b}(y - bS)] - \phi(y)) \right\} = 0. \quad (8.3)$$

If $\lambda = 0$ this is a pure diffusion model, and existence and uniqueness of a smooth solution is to some extent addressed in [7]. When $\lambda > 0$, existence and uniqueness of a solution for the case with $\sigma_P^2 = 0$ and $F$ having a bounded density has been proved through the works [25, 27, 30, 58].

Before discussing specific results, note first that if $r_f > 0$ and $0 \in \mathcal{B}$, with $y > y_0 = (\eta - 1)\frac{p}{r}$, using full reinsurance and having all the capital invested in the risk free asset, investment income is larger than premium loss due to reinsurance. Therefore, for $y > y_0$, $a^*(y) = b^*(y) = 0$ and $\psi^{a^*, b^*}(y) = 0$. This argument is not valid when $r_f = 0$.

Unless stated otherwise, in the following the set $\mathcal{A} = (-\infty, \infty)$.

The first results pertaining directly to this problem were obtained for the pure diffusion case and no reinsurance, i.e. $\lambda = 0$ and $\mathcal{B} = \{1\}$, in [7]. When $r_f = 0$ he proved that $A_t^* = A_0 > 0$, so that it is optimal to invest a fixed amount in the risky asset. When $r_f > 0$, the solution is more complicated, but $\lim_{y\to\infty} A^*(y) = 0$. Also, when $r_f = 0$, $\lim_{y\to\infty} e^{\kappa^* y}\psi^{a^*}(y) = \gamma$ for some known $\kappa^* > 0$ and $\gamma > 0$, while for $r_f > 0$, $\lim_{y\to\infty} e^{\kappa y}\psi^{a^*}(y) = 0$ for all $\kappa > 0$. Thus the asymptotics differ markedly from the case without investment control reported in Section 5, where we saw that the ruin probability goes to zero at a power rate only. The reason is of course that it is optimal to invest only a fixed amount in the risky asset, not a fixed proportion. Further examples for the diffusion model, including some that allow for reinsurance, can be found in [53].

From now on it is assumed that $\lambda > 0$ and that $\sigma_P = 0$, and also that $F$ has a bounded density so that (8.3) has a unique solution.

In the light tailed case, i.e. when $M_S(\kappa) = E[e^{\kappa S}] < \infty$ for some $\kappa > 0$, the results do not differ very much from those in the diffusion case. Although $A_t^*$ is not a constant, when $r_f = 0$ and $\mathcal{B} = \{1\}$, it is proved in [21, 31] that $\lim_{y\to\infty} A^*(y) = A_0 > 0$ and also that

$$\lim_{y\to\infty} e^{\kappa^* y}\psi^{a^*}(y) = \gamma > 0$$

where $\kappa^*$ is the positive solution of

$$\lambda(M_S(\kappa) - 1) - p\kappa = \frac{1}{2}\frac{r}{\sigma_R^2}.$$

When $r_f > 0$ we would expect that $\lim_{y\to\infty} A^*(y) = 0$ as in the diffusion case, but this has not been proved. However, numerical examples given in [39] indicate that this conjecture holds.

The heavy tailed case is somewhat more complicated as the results will vary between subclasses. When $F \in \mathcal{R}_{-\alpha}$ and $\mathcal{B} = \{1\}$, it was proved in [25] that

$$\lim_{y\to\infty} a^*(y) = \frac{r - r_f}{\sigma_R}\frac{1}{1+\alpha},$$
$$\psi^{a^*}(y) \sim \gamma \bar{F}(y)$$

for some known constant $\gamma$. Comparing with the asymptotics of Section 5, it is seen that controlling the investment rate only results in a better convergence rate when investment risk exceeds insurance risk. In that case the control reduces the investment risk so that it is dominated by the insurance risk. Still with $\mathcal{B} = \{1\}$, the bigger class $\mathcal{S}^* \subset \mathcal{S}$ defined by

$$\lim_{y\to\infty}\int_0^y \frac{\bar{F}(y-x)\bar{F}(x)}{\bar{F}(y)}dx = 2E[S]$$

has been studied in [60] when $r_f = 0$ and in [27] when $r_f > 0$. This class is rather big, containing $\mathcal{R}_{-\alpha}$ for $\alpha > 1$, the lognormal distribution and the heavy tailed

Weibull distribution. For this reason, the asymptotics varies within subclasses of $\mathcal{S}^*$, and again whether $r_f = 0$ or not. For example, with $\bar{F}(x) = e^{-\sqrt{x}}$, i.e. heavy tailed Weibull, it is shown in [27] that

$$\frac{\psi^{a^*}(y)}{\psi_0^{a^*}(y)} \sim \frac{\gamma}{\sqrt{y}},$$

for some $\gamma > 0$. Here $\psi_0^{a^*}(y)$ is the minimum ruin probability when $r_f = 0$ and $\psi^{a^*}(y)$ the same when $r_f > 0$.

Finally, when $\mathcal{B} = [0,1]$ and $\bar{F}(x) > ce^{-x^\varepsilon}$ for some $c > 0$ and $0 < \varepsilon < \frac{1}{2}$, it is proved in [59] that for $r_f = 0$ there is a $\gamma > 0$ and a $\kappa^* > 0$ so that

$$\lim_{y \to \infty} e^{\kappa^* y} \psi^{a^*, b^*}(y) = \gamma$$

and

$$\begin{aligned}
\lim_{y \to \infty} A^*(y) &= A_0 > 0, \\
\lim_{y \to \infty} b^*(y) &= 0.
\end{aligned}$$

The reason we obtain an exponential rate of $\psi^{a^*, b^*}(y)$ is that a smaller and smaller fraction of the heavy losses is retained as $Y_t^{a^*, b^*}$ increases.

For more about the problem discussed in this section, the reader can consult [29, 61]. See also [56] for a somewhat different approach.

## References

[1] Albrecher, H., Teugels, J.L. and Tichy, R.F. (2001). On gamma series expansion for the time-dependent probability of collective ruin. *Insurance, Mathematics & Economics*, **29**, 345–355. MR1874629

[2] Albrecher, H., Constantinescu, C. and Thomann, E. (2008). Sparre Andersen models with risky investments. Working paper.

[3] Asmussen, S. (1998). Subexponential asymptotic for stochastic processes: extremal behaviour, stationary distributions and first passage probabilities. *The Annals of Applied Probability*, **8**, 354–374. MR1624933

[4] Asmussen, S. (2000). Ruin Probabilities, World Scientific. MR1794582

[5] Asmussen, S. and Nielsen, H.M. (1995). Ruin probabilities via local adjustment coefficients. *Journal of Applied Probability*, **32**, 736–755. MR1344073

[6] Brekelmans, R. and De Waegenaere, A. (2001). Approximating the finite-time ruin probability under interest force. *Insurance, Mathematics & Economics*, **29**, 217–229. MR1865982

[7] Browne, S. (1995). Optimal investment policies for a firm with a random risk process: Exponential utility and minimizing the probability of ruin. *Mathematics of Operations Research*, **20**, 937–958. MR1378114

[8] Cai, J. and Dickson, D. (2003). Upper bounds for ultimate ruin probabilities in the Sparre Andersen model with interest. *Insurance, Mathematics & Economics*, **32**, 61–71. MR1958769

[9] Cai, J. (2004). Ruin probabilities and penalty functions with stochastic rates of interest. *Stochastic Processes and their Applications*, **112**, 53–78. MR2062567

[10] Cai, J. and Yang, H. (2005). Ruin in the perturbed compound Poisson risk process under interest force. *Advances in Applied Probability*, **37**, 819–835. MR2156562

[11] Cai, J. and Xu, C. (2006). On the decomposition of the ruin probability for a jump-diffusion surplus process compounded by a geometric Brownian motion. *North American Actuarial Journal*, **10**, 120–132. MR2328640

[12] Cai, J. (2007). On the time value of absolute ruin with debit interest. *Advances in Applied Probability*, **39**, 343–359. MR2341577

[13] Cardoso, R.M.R. and Waters, H.R. (2003). Recursive calculation of finite time ruin probabilities under interest force. *Insurance, Mathematics & Economics*, **33**, 659–676. MR2021240

[14] Chen, Y. and Su, C. (2006). Finite time ruin probability with heavy-tailed insurance and financial risks. *Statistics & Probability Letters*, **76**, 1812–1820. MR2274145

[15] Chen, Y. and Ng, K.W. (2007). The ruin probability of the renewal model with constant interest force and negatively dependent heavy-tailed claims. *Insurance, Mathematics & Economics*, **40**, 415–423. MR2310980

[16] Collamore, J.F. (2008). Random recurrence equations and ruin in a Markov-dependent stochastic economic environment. To appear in *The Annals of Applied Probability*.

[17] de Kok, T.G. (2003) Ruin probabilities with compounding assets for discrete time finite horizon problems, independent period claim sizes and general premium structure. *Insurance, Mathematics & Economics*, **33**, 645–658. MR2021239

[18] Delbaen, F. and Haezendonck J. (1987). Classical risk theory in an economic environment. *Insurance, Mathematics & Economics*, **6**, 85–116. MR0896414

[19] Embrechts, P. and Schmidli, H. (1994). Ruin estimation for a general insurance risk model. *Advances in Applied Probability*, **26**, 404–422. MR1272719

[20] Frolova, A.G., Kabanov, Y. and Pergamenshchikov, S.M. (2002). In the insurance business risky investments are dangerous. *Finance and Stochastics*, **6**, 227–235. MR1897960

[21] Gaier, J., Grandits, P. and Schachermayer, W. (2003). Asymptotic ruin probabilities and optimal investment. *The Annals of Applied Probability*, **13**, 1054–1076. MR1994044

[22] Gerber, H.U. (1971). Der Einfluss von Zins auf die Ruinwahrscheinlichkeit. *Mitteilungen der Schweizerischer Vereinigung der Versicherungsmatematiker*, 63–70.

[23] Gerber, H.U. and Shiu, E.S.W. (1998). On the time value of ruin, *North American Actuarial Journal*, **2**, 48–78. MR1988433

[24] Gerber, H. and Yang, H. (2007). Absolute ruin probabilities in a jump diffusion risk model with investment. *North American Actuarial Journal*, **11**, 159–169. MR2393866

[25] Gaier, J. and Grandits, P. (2004). Ruin probabilities and investment un-

der interest force in the presence of regularly varying tails. *Scandinavian Actuarial Journal*, 256–278. MR2081815
- [26] Grandits, P. (2004). A Karamata-type theorem and ruin probabilities for an insurer investing proportionally in the stock market. *Insurance, Mathematics & Economics*, **34**, 297–305. MR2053791
- [27] Grandits, P. (2005). Minimal ruin probabilities and investment under interest force for a class of subexponential distributions. *Scandinavian Actuarial Journal*, 401–416. MR2202583
- [28] Harrison, J.M. (1977). Ruin problems with compounding assets. *Stochastic Processes and their Applications*, **5**, 67–79. MR0423736
- [29] Hipp, C. (2004). Stochastic control with application in insurance. *Stochastic methods in finance, Lecture Notes in Math.*, **1856**, 127–164, Springer, Berlin. MR2113722
- [30] Hipp, C. and Plum, M. (2000). Optimal investment for insurers. *Insurance, Mathematics & Economics*, **27**, 215–228. MR1801604
- [31] Hipp, C. and Schmidli, H. (2004). Asymptotics of ruin probabilities for controlled risk processes in the small claim case. *Scandinavian Actuarial Journal*, 321–335. MR2096062
- [32] Jiang, T. and Yan, H-F. (2006). The finite-time ruin probability for the jump-diffusion model with constant interest force. *Acta Mathematicae Applicatae Sinica*, **22**, 171–176. MR2191727
- [33] Kalashnikov, V. and Konstantinides, D. (2000). Ruin under interest force and subexponential claims: a simple treatment. *Insurance, Mathematics & Economics*, **27**, 145–149. MR1796976
- [34] Kalashnikov, V. and Norberg, R. (2002). Power tailed ruin probabilities in the presence of risky investments. *Stochastic Processes and their Applications*, **98**, 211–228. MR1887534
- [35] Klüppelberg, C. and Stadtmuller, U. (1998). Ruin probabilities in the presence of heavy-tails and interest rates. *Scandinavian Actuarial Journal*, 49–58. MR1626664
- [36] Knessl, C. and Peters, C. (1994). Exact and asymptotic solutions for the time-dependent problem of collective ruin I. *SIAM Journal of Applied Mathematics*, **54**, 1745–1767. MR1301280
- [37] Konstantinides, D.G., Tang, Q.H. and Tsitsiashvili, G.S. (2002). Estimates for the ruin probability in the classical risk model with constant interest force in the presence of heavy tails. *Insurance, Mathematics & Economics*, **31**, 447–460. MR1945543
- [38] Konstantinides, D.G., Tang, Q.H. and Tsitsiashvili, G.S. (2004). Two-sided bounds for ruin probability under constant interest force. *Journal of Mathematical Sciences*, **123**, 3824–3833. MR2093829
- [39] Liu, C-S. and Yang, H. (2004). Optimal investment for an insurer to minimize its probability of ruin. *North American Actuarial Journal*, **8**, 11–31. MR2064431
- [40] Lundberg, F. (1903). *Approximerad Framstilling av Sannolikhetsfunktionen II. Återforsäkring av Kollektivrisker*, Almquist & Wiksell, Uppsala.
- [41] Ma, J. and Sun, X. (2003). Ruin probabilities for insurance models involving

investments. *Scandinavian Actuarial Journal*, 217–237. MR1996926
[42] Norberg, R. (1999). Ruin problems with assets and liabilities of diffusion type. *Stochastic Processes and their Applications*, **81**, 255–269. MR1694553
[43] Nyrhinen, H. (1999). On the ruin probabilities in a general economic environment. *Stochastic Processes and their Applications*, **89**, 319–330. MR1708212
[44] Nyrhinen, H. (2001). Finite and infinite time ruin probabilities in a stochastic economic environment. *Stochastic Processes and their Applications*, **92**, 265–285. MR1817589
[45] Paulsen, J. (1993). Risk theory in a stochastic economic environment. *Stochastic Processes and their Applications*, **46**, 327–361. MR1226415
[46] Paulsen, J. (1998). Sharp conditions for certain ruin in a risk process with stochastic return on investments. *Stochastic Processes and their Applications*, **75**, 135–148. MR1629034
[47] Paulsen, J. (1998). Ruin theory with compounding assets - a survey. *Insurance, Mathematics & Economics*, **22**, 3–16. MR1625827
[48] Paulsen, J. (2002). On Cramér-like asymptotics for risk processes with stochastic return on investments. *The Annals of Applied Probability*, **12**, 1247–1260. MR1936592
[49] Paulsen, J. and Gjessing, H.K. (1997). Ruin theory with stochastic return on investments. *Advances in Applied Probability*, **29**, 965–985. MR1484776
[50] Paulsen, J. and Rasmussen, B.N. (2003). Simulating ruin probabilities for a class of semimartingales by importance sampling methods. *Scandinavian Actuarial Journal*, 178–216. MR1996925
[51] Paulsen, J., Kasozi, J., and Steigen, A. (2005). A numerical method to find the probability of ultimate ruin in the classical risk model with stochastic return on investments. *Insurance, Mathematics & Economics*, **36**, 399–420. MR2152852
[52] Pergamenshchikov, S. and Zeitouny, O. (2006). Ruin probability in the presence of risky investments. *Stochastic Processes and their Applications*, **116**, 267–278. MR2197977
[53] Promislow, S.D. and Young, V.R. (2005). Minimizing the probability of ruin when claims follow Brownian motion with drift. *North American Actuarial Journal*, **9**, 109–128. MR2200024
[54] Rolski, T., Schmidli, H., Schmidt, V. and Teugels, J.L. (1999). *Stochastic Processes for Insurance and Finance*, Wiley, Chichester. MR1680267
[55] Ruohonen, M. (1980). On the probability of ruin of risk processes approximated by diffusion processes. *Scandinavian Actuarial Journal*, 113–120. MR0578451
[56] Schäl, M. (2005). Control of ruin probabilities by discrete-time investments. *Mathematical Methods of Operations Research*, **62**, 141–158. MR2226972
[57] Schmidli, H. (1994). Risk theory in an economic environment and Markov processes. *Mitteilungen der Vereinigung der Versicherungsmatematiker*, 51–70. MR1293083
[58] Schmidli, H. (2002). On minimizing the ruin probability by investment and reinsurance. *The Annals of Applied Probability*, **12**, 890–907. MR1925444

[59] Schmidli, H. (2004). Asymptotics of ruin probabilities for risk processes under optimal reinsurance and investment policies: The large claim case. *Queueing Systems*, **46**, 149–157. MR2072280

[60] Schmidli, H. (2005). On optimal investment and subexponential claims. *Insurance, Mathematics & Economics*, **36**, 25–35. MR2122663

[61] Schmidli, H. (2008). *Stochastic Control in Insurance*, Springer. MR2371646

[62] Segerdahl, C.O. (1942). Über einige risikotheoretische Fragestellungen. *Skandinavisk Aktuartidsskrift*, **25**, 43–83. MR0017902

[63] Sundt, B. and Teugels, J.L. (1995). Ruin estimates under interest force. *Insurance, Mathematics & Economics*, **16**, 7–22. MR1342906

[64] Tang, Q. and Tsitsiashvili, G. (2003). Precise estimates for the ruin probability in finite horizon in a discrete-time model with heavy-tailed insurance and financial risks. *Stochastic Processes and their Applications*, **108**, 299–325. MR2019056

[65] Tang, Q. and Tsitsiashvili, G. (2004). Finite- and infinite ruin probabilities in the presence of stochastic returns on investments. *Advances in Applied Probability*, **36**, 1278–1299. MR2119864

[66] Tang, Q. (2004). The ruin probability of a discrete time risk model under constant interest rate with heavy tails. *Scandinavian Actuarial Journal*, 229–240. MR2064607

[67] Tang, Q. (2005). Asymptotic ruin probabilities of the renewal model with constant interest force and regular variation. *Scandinavian Actuarial Journal*, 1–5. MR2118521

[68] Tang, Q. (2005). The finite time ruin probability of the compound Poisson model with constant interest force. *Journal of Applied Probability*, **42**, 608–619. MR2157508

[69] Wang, G. and Wu, R. (2001). Distributions for the risk process with a stochastic return on investments. *Stochastic Processes and their Applications*, **95**, 329–341. MR1854031

[70] Wang, G., Yang, H. and Wang, H. (2004). On the distribution of surplus immediately after ruin under interest force and subexponential claims. *Insurance, Mathematics & Economics*, **35**, 703–714. MR2106144

[71] Wang, G. and Wu, R. (2008). The expected discounted penalty function for the perturbed compound Poisson risk process with constant interest. *Insurance, Mathematics & Economics*, **42**, 59–64. MR2392069

[72] Wu, R., Wang, G. and Zhang, C. (2005). On a joint distribution for the risk process with constant interest force. *Insurance, Mathematics & Economics*, **36**, 365–374. MR2152850

[73] Yang, H. and Zhang, L. (2001). The joint distribution of surplus immediately before ruin and the deficit at ruin under interest force. *North American Actuarial Journal*, **5**, 92–103. MR1989761

[74] Yang, H. and Zhang, L. (2006). Ruin problems for a discrete time risk model with random interest rate. *Mathematical Methods of Operations Research*, **63**, 287–299. MR2264750

[75] Yuen, K.C., Wang, G. and Ng, K.W. (2004). Ruin probabilities for a risk process with stochastic return on investments. *Stochastic Processes and*

*their Applications*, **110**, 259–274. MR2040968
[76] Yuen, K.C. and Wang, G. (2005). Some ruin problems for a risk process with stochastic interest. *North American Actuarial Journal*, **9**, 129–142. MR2200025
[77] Yuen, K.C., Wang, G. and Wu, R. (2006). On the renewal risk process with stochastic interest. *Stochastic Processes and their Applications*, **116**, 1496–1510. MR2260745
[78] Zhu, J. and Yang, H. (2008). Estimates for the absolute ruin probability in the compound Poisson risk model with credit and debit interest. *Advances in Applied Probability*, **40**, 818–830.